\input psfig

\documentstyle{article}

\begin{document}

\vspace{1 cm}
\Large
\title{\bf On the reliability of mean-field methods in polymer statistical 
mechanics}
\vspace{1.8 cm}
\large
\author{Stefan Tsonchev and Rob D. Coalson \\
Department of Chemistry, University of Pittsburgh, Pittsburgh, PA 15260
\and
Shyh-Shi Chern and Anthony Duncan \\
Department of Physics, University of Pittsburgh, Pittsburgh, PA 15260}
\date{}
\maketitle

\begin{abstract}
The reliability of the mean-field approach to polymer statistical mechanics 
is investigated by comparing results from a recently developed lattice 
mean-field theory (LMFT) method to statistically exact results from two 
independent numerical Monte Carlo simulations for the problems of a 
polymer chain 
moving in a spherical cavity and a polymer chain partitioning between 
two confining spheres of different radii. It is shown that in some cases 
the agreement between the LMFT and the simulation results is excellent, while 
in others, such as the case of strongly fluctuating monomer repulsion fields, 
the LMFT results agree with the simulations only qualitatively. Various 
approximations of the LMFT method are systematically estimated, 
and the quantitative discrepancy between the two sets of results is explained 
with the diminished accuracy of the saddle-point approximation, implicit in 
the mean-field method, in the case of strongly fluctuating fields.
\end{abstract}

\newpage
\section{Introduction}
Recently, there has been renewed interest \cite{Asher,Chern,Ts,TCD} in the
venerable problem of polymer partitioning \cite{Cas,Mutu}. This phenomenon 
is of interest both from a
theoretical as well as a practical perspective. Macromolecular separation 
techniques, 
such as size exclusion chromatography, gel electrophoresis, filtration, 
membrane separation, etc. \cite{Rod} all have their basis in polymer 
partitioning---the property of macromolecular chains to distribute themselves 
in a network of random obstacles according to their molecular weight/length 
or electric charge. These phenomena are also of fundamental interest for 
understanding the properties of macromolecules and complex fluids.

In some recent work \cite{TCD2,TCD} we developed a lattice field theory 
approach
to the statistical mechanics of polymers in solution. This approach relied on 
several approximations: the treatment was at the mean-field level, which 
amounts to a  saddle-point approximation to the system's partition function; 
also the problem was treated on a lattice, converting a continuous three 
dimensional (3D) problem to a discrete, finite matrix problem, and, finally, 
 the polymer chain was assumed to be long enough so that it was 
approximated as a continuous object, ignoring the discrete character of the 
building monomers.

The goal of this paper is to investigate the range of validity of these 
approximations. In order to do so we need to make a comparison with 
an exact solution of the problem, which can be obtained in the case of 
neutral polymers by numerical Monte Carlo simulations. We are mostly 
interested in the 
reliability of the mean-field approach and its implicit saddle-point 
approximation, and the cases in which it can be applied with a high level of 
accuracy. 

In Section 2 of the paper we generalize our previous treatment to the case 
of a soft short-range interaction potential  between the monomers in the 
polymer chain, which is needed for the direct comparison with the simulation 
results. In this section we also discuss the other two types of errors, namely
the lattice discretization error and the error due to the approximation of 
the chain as a continuous object---we shall call it the ``chain continuity 
error''---and the need 
for their elimination in order to estimate the reliability of the mean-field 
approach per se. In Section 3 we show how the chain continuity error 
can be  eliminated by a transfer matrix representation of the partition 
function, and
discuss the effects of this error for the problem of a polymer chain in a 
spherical cavity. Here also we show how the transfer matrix approach can be 
extended to the more general case of a 3D lattice. In Section 4 we present 
results for the monomer distribution of a polymer chain in a sphere from
 the transfer 
matrix approach and compare them with those for the case of a continuous 
chain, and with the exact simulation results, showing a good agreement 
between the three approaches.
In Section 5 we describe the procedures used in two different numerical 
simulation methods, and  present the results obtained by these two
 independent methods compared with the results obtained by the 
mean-field method using the transfer matrix approach, for the problem of a 
polymer chain confined to move within the region of two connected spheres. 
The results show that for certain range of parameters there is a substantial 
deviation between the mean-field approach and the simulations, which is 
clearly due to the large fluctuations in the monomer repulsion field. We are
presently working on a method that would allow us to overcome this problem 
by applying the mean-field approach separately in different regions where it 
is accurate, and ``welding'' the results for the polymer partition function 
of the separate regions to obtain an accurate description of the more 
complicated physical situation. In Section 6 we outline 
possible applications of the methods and conclude.

\newpage
\section{A strategy for studying the accuracy of the mean-field 
approximation in polymer problems}
  In order to study systematically the level of accuracy of the mean-field 
approach
 to polymer statistical mechanics introduced in some recent work of the 
present
 authors \cite{TCD2,TCD}, it will be convenient to introduce a  
slightly
 generalized Hamiltonian which allows for a soft Yukawa type repulsive 
potential
 between the monomers of the polymer chain \cite{CWDB}. In such a 
model
 (absent long-range electrostatic interactions, which are not treated in 
this paper)
 an essentially exact (up to statistical errors) solution of the statistical 
mechanics
 is available via standard Monte Carlo simulation methods, while, as we 
shall see below,
 the mean-field Schr\"odinger approach of \cite{TCD2,TCD} can be generalized 
to such a 
 model, allowing a direct comparison and evaluation of the error incurred by 
the 
 saddle-point approximation implicit in the mean-field equations. 

  The partition sum for a Gaussian polymer chain of $M$ monomers 
with a repulsive Yukawa potential
  acting between any pair of monomers on the chain and a general exclusion
 potential $V_{\rm excl}(\vec{r})$ (which can be used to exclude monomers 
from certain regions of space) can be written

\begin{eqnarray}
    Z_{{\rm pol}}&=&\int\prod_{s} d\vec{x}_{s}e^{-\frac{3}{2a_{p}^{2}}\sum_{s=0}^{M-1} (\vec{x}_{s+1}-\vec{x}_{s})^2
 -\frac{\lambda}{2}\sum_{s,s^{\prime}}V_{\rm Yuk}(\vec{x}_{s}-\vec{x}_{s^{\prime}})
 -\sum_{s}V_{\rm excl}(\vec{x}_{s})}.
\label{Z}
\end{eqnarray}
Here $s,s^{\prime}$ are integers labelling the monomer location on the chain 
and the Yukawa repulsion between two monomers with coordinate separation 
$\vec{r}$ is
\begin{equation}
  V_{\rm Yuk}(\vec{r}) = \frac{e^{-|\vec{r}|/\sqrt{\gamma}}}{4\pi\gamma |\vec{r}|}.
\label{Vyuk}
\end{equation}
 The unconventional choice of $\sqrt{\gamma}$ for the range of the potential 
will simplify
 the algebra later on (in particular, the attentive reader  will 
recognize that $V_{\rm Yuk}$ is exactly the Green's function of
 the differential operator $1-\gamma\Delta$). The partition function in  (1) 
involves a positive Boltzmann measure with
 short range interactions and is therefore susceptible to a standard Monte 
Carlo simulation.
 We have found it convenient to perform the simulation by updating the 
monomer locations
 with a heat-bath algorithm for the Gaussian part of the Hamiltonian in (1), 
followed by
 a Metropolis accept/reject step to include the effects of the Yukawa and 
exclusion potentials $V_{\rm Yuk},V_{\rm excl}$ (cf. Section 4 for more 
details).

    In the mean-field approach to polymer equilibrium statistical mechanics 
developed
 in our recent work \cite{TCD2,TCD}, we have generally assumed that the 
polymer chains are
 sufficiently long that the discrete sums over the monomer index $s$ in (1) 
may be
 replaced by a continuous integral over a dimensionless variable $s$, also 
running from
 0 to $M\equiv$ total number of monomers in the chain. From the point of 
view of the
 original model (1) (for which we shall have essentially 
exact---up to controllable statistical error---results via simulation) this
 is an approximation, which we have termed the 
``chain continuity approximation" (to distinguish it from the lattice 
discretization approximation, for example,
 in which fields on continuous 3-space are replaced by fields defined on a 
3D spatial lattice).  In this approximation, (1) becomes a path integral
\begin{equation}
    Z_{{\rm pol}}=\int D\vec{x}(s)e^{-\frac{3}{2a_{p}^{2}}\int_{0}^{M}ds \dot{\vec{x}}^{2}(s)
 -\frac{\lambda}{2}\int d\vec{r}d\vec{r}\,'j(\vec{r})V_{\rm Yuk}(\vec{r}-\vec{r}\,')j(\vec{r}\,') -\int _{0}^{M}dsV_{\rm excl}(\vec{x}(s))},
\label{Zpi}
\end{equation}
with
\begin{equation}
   j(\vec{r}) \equiv \int ds \delta(\vec{r}-\vec{x}(s)),
\label{j}
\end{equation}
 where we have introduced a line distribution $j(\vec{r})$ with support 
along the line tracing 
 through the monomers. To establish contact  with the models studied in our 
previous work
 \cite{TCD2,TCD}, we linearize the dependence on $j$ by introducing an 
auxiliary field $\omega$
 via a Hubbard-Stratonovich transformation. The path integral 
(3) then becomes
\begin{equation}
     Z_{{\rm pol}}=\int D\vec{x}(s)D\omega(\vec{r})e^{-\frac{3}{2a_{p}^{2}}\int_{0}^{M}ds \dot{\vec{x}}^{2}(s)
 -\frac{\lambda}{2}\int d\vec{r}\omega(\vec{r})(1-\gamma\Delta)\omega(\vec{r})
 -i\lambda\int d\vec{r}j(\vec{r})\omega(\vec{r})}.
\label{Zpi2}
\end{equation}
 With the range $\sqrt{\gamma}$ set to zero, we recover the simpler model 
studied in our earlier
 mean-field papers \cite{TCD2,TCD} for the special case of uncharged 
polymer chains. Here we
 have specialized to the case of a single particle potential $V_{\rm excl}$ 
in (\ref{Zpi}) which is
 either zero or $+\infty$, and the sole effect of which is to restrict the 
fields in the path
integral (\ref{Zpi2}) to have support in the region where $V_{\rm excl}=0$.
 The path 
 integral over $\vec{x}(s)$ can be replaced in the standard way by an 
equivalent
 Schr\"odinger problem for the imaginary time evolution (from 0 to $M$) 
of a particle
 of mass 3/$a_{p}^{2}$ in an imaginary potential $i\lambda\omega(\vec{r})$. 
Thus
\begin{equation}
 Z_{\rm{pol}}=\int D\omega(\vec{r})e^{
-\frac{\lambda}{2}\int\omega(\vec{r})(1-\gamma\Delta)\omega(\vec{r})
}Z_{{\rm Schr}}(\omega),
\label{Zpi3}
\end{equation}
with 
\begin{equation}
Z_{{\rm Schr}}(\omega)\equiv \int D\vec{x}(s)e^{-\frac{3}{2a_{p}^{2}}\int_{0}^{M}ds \dot{\vec{x}}^{2}(
s)-i\lambda\int ds\omega(\vec{x}(s))}.
\label{Zschr}
\end{equation}
 The mean-field approximation to the system (\ref{Zpi3}--\ref{Zschr}) 
amounts to rerouting the functional integral
 over $\omega$ through a complex saddle point at $\omega = -i\omega_{c}$, 
where $\omega_{c}$ is a {\em real} function, at which point the evaluation 
of the Schr\"odinger
 amplitude $Z_{\rm Schr}$ in (\ref{Zschr}) reduces to a conventional 3D 
quantum mechanical
 evolution (in imaginary time) of a system subject to the Hamiltonian
\begin{equation}
   H \equiv -\frac{a_{p}^{2}}{6}\Delta+\lambda\omega_{c}(\vec{r}).
\label{H}
\end{equation}
 The hermitian Hamiltonian (\ref{H}) has a complete set of normalized 
eigenfunctions $\Psi_{n}(\vec{r})$
 with corresponding eigenvalues $E_{n}$, and the evolution amplitude 
$Z_{\rm Schr}$ appearing in (\ref{Zschr}) is just 
\begin{equation}
  Z_{\rm Schr}=\int d\vec{x}_{0}d\vec{x}_{M}<\vec{x}_{0}|e^{-MH}|\vec{x}_{M}>=\int dx_{0}dx_{M}\sum_{n}\Psi_{n}(x_{0})\Psi_{n}(x_{M})e^{-ME_{n}}.
\label{Zschr2}
\end{equation}
  It has been conventional to assume ground state dominance in evaluating  
$Z_{\rm Schr}$,
 but  this condition is frequently violated in the
 systems studied below \cite{Ts,TCD}.  In a recent paper, we have presented 
the general solution
 \cite{TCD} which makes no assumption of ground state dominance. Defining
 $Z_{\rm Schr}\equiv e^{F_{\rm pol}}$, and
\begin{equation}
 A_{n} \equiv \int d\vec{r}\Psi_{n}(\vec{r}),
\label{An}
\end{equation}
we have
\begin{equation}
  F_{\rm pol} = \ln{\sum_{n} A_{n}^{2}e^{-ME_{n}}}.
\label{F}
\end{equation}
The saddle point of the functional integral (\ref{Zpi3}) is then found by 
minimizing the functional
\begin{equation}
 F = \frac{\lambda}{2}\int d\vec{r} \omega_{c}(\vec{r})(1-\gamma\Delta)\omega_{c}(\vec{r})
 +F_{\rm pol}(\omega_{c}).
\label{Ftot}
\end{equation}
Varying this functional as in our previous work \cite{TCD}, we obtain:
\begin{equation}
\frac{{\delta}F_{pol}}{\delta\omega_{c}(\vec{r})}=\lambda\rho(\vec{r}),
\label{varF/varomega}
\end{equation}
where
\begin{equation}
\rho(\vec{r})=-\frac{\sum_{n,m}\frac{A_{n}\Psi_{n}A_{m}\Psi_{m}}{E_{n}-E_{m}}\left(e^{-ME_{n}}-e^{-ME_{m}}\right)}{\sum_{n}A_{n}^{2}e^{-ME_{n}}}. 
\label{rho}
\end{equation}
Therefore,
\begin{equation}
   \frac{1}{\lambda} \frac{\delta F}{\delta\omega_{c}(\vec{r})} 
=(1-\gamma\Delta)\omega_{c}(\vec{r})-\rho(\vec{r})=0 \, .
\label{fe}
\end{equation}
Equation (\ref{fe}) implies
\begin{equation}
\omega_{c}(\vec{r})\equiv \hat{\rho}(\vec{r})=\int V_{Yuk}(\vec{r}-\vec{r}\,')\rho(\vec{r}\,')d\vec{r}\,',
\label{omegac}
\end{equation}
 so that the saddle-point value of the auxiliary field $\omega(\vec{r})$ can 
be thought of either as a smeared version of the monomer density $\rho$ or 
as the total repulsive potential at $\vec{r}$ due to all monomers. 
We may therefore 
eliminate the auxiliary field altogether in
(\ref{H}), leading to the following mean-field equation describing the 
equilibrium properties of the polymer chain:
\begin{equation}
  \frac{a_{p}^{2}}{6}\vec{\nabla}^{2}\Psi_{n}(\vec{r})=\lambda 
\hat{\rho}(\vec{r})\Psi_{n}(\vec{r})-E_{n}\Psi_{n}(\vec{r}).
\label{nlse}
\end{equation}
This equation is a generalization of Eq. (22) in our previous work \cite{TCD}
for the Yukawa type interaction potential between monomers. It is a 
continuous mean-field  nonlinear Schr\"odinger type equation which in
 general must be solved numerically by putting the system on a discrete 
spatial lattice. This
  introduces (in addition to the errors incurred by making the chain 
continuous and the saddle-point
 approximation of (\ref{Zpi2})) a third source of error in comparing the 
results with the exact
 solution (via Monte Carlo) of the original system (\ref{Z}). The 
determination of the qualitative effects and relative 
 size of  all three sources of error in some systems of interest is the 
primary focus of this paper.
 In particular, it will be important to understand the size of the chain 
continuity error and lattice discretization
 effects in determining the extent to which the mean-field approximation 
per se  fails or succeeds in a specific polymer problem.

\newpage
\section{Discrete vs continuous chains: a transfer matrix approach}
  In this section we estimate the size of the error incurred by replacing a 
discrete chain of
 $M$+1 monomers in (\ref{Z}) by a continuous imaginary-time Schr\"odinger 
evolution as in
 (\ref{Zschr}).  For simplicity, we first study this issue with $\lambda$ = 0 
(no intermonomer repulsion)
 for a polymer chain confined to a spherical region of radius $R$. In this 
simple case
 the main qualitative effect of this replacement is to force the monomer 
density to vanish
 exactly at the boundary of the sphere, whereas there is a finite average 
density of
 monomers at the boundary in the case of the discrete chain \cite{Gro}. 
It turns out that a transfer
 matrix formalism allows a simple analytic description of  the discrete 
chain case, which
 enables us to derive an explicit formula for the size of this ``wall effect".

  Referring to the partition sum (\ref{Z}), introduce a transfer matrix 
$T(\vec{x},\vec{x}\,')$ as follows:
\begin{equation}
  T(\vec{x},\vec{x}\,')\equiv \Theta_{S}(\vec{x})\Theta_{S}(\vec{x}\,')e^{-\alpha(\vec{x}-\vec{x}\,')^{2}},\;\;\;\alpha\equiv
\frac{3}{2a_{p}^{2}},
\label{tdef}
\end{equation}
 where $\Theta_{S}(\vec{x})$ is the Heaviside theta-function with support 
inside the sphere of
 radius $R$. Viewing the transfer-matrix as a linear operator kernel, the 
partition function (\ref{Z}) can then be written simply as:
\begin{equation}
 Z_{\rm pol} = \int d\vec{x}_{0}d\vec{x}_{M} T^{M}(\vec{x}_{0},\vec{x}_{M}).
\label{Zpi4}
\end{equation}
On the other hand, we can define a Hamiltonian $H_{\rm d}$ for the discrete 
chain by the usual connection,
 $T=e^{-H_{\rm d}}$, so that 
\begin{equation}
 Z_{\rm pol} = \int d\vec{x}_{0}d\vec{x}_{M} <\vec{x}_{0}|e^{-MH_{\rm d}}|\vec{x}_{M}>.
\label{Zpi5}
\end{equation}
Of course, the operator $H$ in (\ref{H}) and (\ref{Zschr2}) and $H_{\rm d}$ 
differ, and it is precisely this
 discrepancy which we have referred to previously  as the 
``chain continuity error". 
 Note that the integrals over initial and final monomer locations in 
Eqs. (\ref{Zschr2}) and (\ref{Zpi5})
 (corresponding to free boundary conditions for an open polymer chain) 
restrict us to
 s-wave (spherically symmetric) eigenfunctions of both $H$ and $H_{\rm d}$. 
In the case of
 the Schr\"odinger operator $H$, these eigenfunctions are the familiar 
spherical Bessel
 functions and the corresponding eigenvalues are related in the usual way 
to the zeroes
 of these radial functions [as $\Psi_{n}(r{=}R)=0$ in the continuous chain 
approximation].
 For the discrete chain operator, we must diagonalize the integral kernel 
$T$ in (\ref{tdef}).
 The eigenvalues $\varepsilon_{n}$ of $T$ are related to the eigenvalues 
$E_{n}$ of $H_{\rm d}$
 by $\varepsilon_{n}=e^{-E_{n}}$, while the eigenfunctions of $T$ and 
$H_{\rm d}$ are the same.
 Restricting ourselves to spherically symmetric eigenfunctions, we need to 
solve the linear integral equation
\begin{equation}
 \int_{r'<R} e^{-\alpha|\vec{r}-\vec{r}\,'|^{2}}\psi(\vec{r}\,')d\vec{r}\,'=\varepsilon\psi(\vec{r}),
\label{inte}
\end{equation}
which reduces after integration over angles to the one dimensional (1D)
 integral equation
\begin{equation}
 \frac{\pi}{\alpha}\int_{0}^{R}dr'\left(e^{-\alpha(r-r^{\prime})^{2}}-e^{-\alpha(r+r^{\prime})^{2}}\right)f(r^{\prime}) = \varepsilon f(r),
\label{inte2}
\end{equation}
where $f(r)\equiv r\psi(r)$. This equation can easily be solved by 
discretization and numerical
 diagonalization, whence one finds the ``wall effect" previously mentioned, 
namely $f(R) \neq 0$ (this effect is also visible in explicit simulations, 
cf. Section 4). We can estimate the
 size of the effect in the limit $a_{p}\ll R$ (which implies 
$\alpha R^{2}\gg 1$ and the
 negligibility of the second exponential for $r \simeq R$ in 
(\ref{inte2})) as follows.
  To leading order in $a_{p}/R$, the radial transfer matrix in 
(\ref{inte2})) is
\begin{equation}
  \int e^{-\alpha(r-r^{\prime})^{2}}f(r^{\prime})dr^{\prime} \simeq \sqrt{\frac{\pi}{\alpha}}\left(1+\frac{1}{4\alpha}\frac{d^{2}}{dr^{2}}\right)f(r).
\label{inte3}
\end{equation}
The ground state corresponds to the eigenvalue 
$-\left(\frac{\pi}{R}\right)^{2}$ of $\frac{d^{2}}{dr^{2}}$,
 so in this limit the $\varepsilon$ eigenvalue of the ground state is 
$\varepsilon_{0}\approx\left(\frac{\pi}{\alpha}\right)^{3/2}
\left(1-\frac{\pi^{2}}{4\alpha R^{2}}\right)$.
Next, writing 
$f(r)\simeq\delta+\sin\left(\pi\frac{r}{R}\right),\;\;r\simeq R$, 
and substituting this ansatz into (\ref{inte3}) at $r=R$, we find
\begin{equation}
 \int e^{-\alpha(r^{\prime}-R)^{2}}\left(\delta+\sin\left(\pi\frac{r^{\prime}}{R}\right)\right)dr^{\prime}\simeq\frac{\varepsilon\alpha}{\pi}\delta.
\label{inte4}
\end{equation}
 The radial integral can be evaluated in the given approximation by 
linearizing the sine function near $r=R$, 
$\sin\left(\frac{\pi r'}{R}\right)\approx \frac{\pi}{R}(R-r')$, 
whereupon we find
\begin{equation}
  \frac{\varepsilon\alpha}{\pi}\delta \simeq \frac{1}{2}\sqrt{\frac{\pi}{\alpha}}\delta+\frac{1}{2\alpha}\frac{\pi}{R}.
\label{intsol}
\end{equation}
 Substituting the value $\varepsilon_{0}$ found above for the ground state, 
we find the size of the wall effect to leading order in $1/\alpha R^{2}$ to be
\begin{equation}
  \delta \simeq \sqrt{\frac{\pi}{\alpha}}\frac{1}{R}=\sqrt{\frac{2\pi}{3}}\frac{a_{p}}{R}.
\label{walle}
\end{equation}
It should be noted that this estimate is conditioned on $f(r)$ being 
approximated by $\sin\left(\pi\frac{r}{R}\right)$, that is, $f(r)$ is 
normalized to be approximately unity at $r=R/2$.
In the next section we shall present numerical results for the single sphere system confirming
 this analytic estimate.

  For more general problems in three dimensions, lacking spherical 
symmetry (such as the two sphere situation studied in Section 5)
 it is still possible to diagonalize the transfer matrix,
 in this case by going to a discrete spatial lattice with lattice spacing 
$a$. Then the transfer matrix (in this case, truly a finite dimensional 
matrix!) becomes
\begin{equation}
  T_{\vec{n}\vec{m}}=e^{-\frac{3a^{2}}{2a_{p}^{2}}|\vec{n}-\vec{m}|^{2}-\frac{1}{2}(V_{\vec{n}}+V_{\vec{m}})},
\label{tm}
\end{equation}
 where $V_{\vec{n}}$ is the value of the single-particle potential 
$V_{\rm excl}$ in (\ref{Z}) at the
 lattice point $\vec{n}$ (recall that we are still taking $\lambda=0$ in 
(\ref{Z})).
 Although in principle a non-sparse matrix, we have found it perfectly 
adequate to cutoff the
 matrix elements of (\ref{tm}) for $|\vec{n}-\vec{m}|>6$, whereupon the 
remaining
 relatively sparse matrix can be diagonalized by standard Lanczos 
techniques. The results
 of calculations employing this transfer matrix, and therefore eliminating 
the chain continuity
error in comparisons with the Monte Carlo simulations of (\ref{Z}), are 
given in the following two sections.

  If the intermonomer repulsion $V_{\rm Yuk}$ is present then the transfer 
matrix formalism
 can still be employed to estimate the chain continuity error, but 
only in the context of
 the mean-field approximation discussed in the previous section. After 
introducing the
 auxiliary field $\omega$ and going to the complex saddle point at 
$\omega= -i\omega_{c}$, we can reinstate a discrete chain by 
using the transfer matrix (\ref{tm}) with $V_{\vec{n}}$ taken equal to the 
value of  $ V_{\rm excl}+\lambda\omega_{c}$ at the lattice point $\vec{n}$, 
where $\omega_{c}$ is again given by the formula (\ref{omegac}) and 
$\rho(\vec{r})$ by (\ref{rho}), with slightly different coefficients 
$A_{d,n}$ defined as
\begin{equation}
A_{d,n}\equiv \int d\vec{r}e^{-(1/2)V(\vec{r})}\Psi_{n}(\vec{r}).
\label{Adn}
\end{equation}

\newpage
\section{Comparison of transfer matrix and simulation results for a 
polymer chain in a spherical cavity}

In this section we consider a polymer chain of 100 monomers, restricted
to move within the volume of a spherical cavity. We have chosen the Kuhn 
length to be unity and a sphere radius equal to ten. First, we discretize the 
integral equation (\ref{inte2}) (neglecting the second exponential) 
on a fine 1D lattice, thus converting it into 
a standard  matrix eigenvalue problem, which we solve to a high precision.
The solution involves only spherically symmetric s-waves.
In Fig. 1 we plot the function $f(r)$ defined in the previous section 
by Eq. (\ref{inte2}), 
which clearly exhibits the ``wall effect'' discussed earlier---it has a 
finite value at the edge of the sphere. The value $f(R)$ in Fig. 1 is 
0.139, which is close to the value calculated from Eq. (\ref{walle}) for 
the given set of parameters, $\delta\approx 0.145$. 

Then we apply the Lanczos procedure described in our earlier work 
\cite{TCD2,TCD} to obtain the eigenvalues and eigenvectors of the 
transfer matrix (\ref{tdef}), and therefore the monomer density. The 
matrix is discretized on a cubic lattice of 44 points on each side, with 
lattice spacing $a=a_{p}/2$, in which the spherical cavity is embedded. The 
resulting radial monomer density is plotted on Fig. 2, where it is compared 
with the exact result from a Monte Carlo heat bath simulation (described 
in the next section), the 
1D transfer matrix result of Eq. (\ref{inte2}), and the result from the 
Schr\"odinger equation approach, Eq. (\ref{nlse}), for $\lambda$ = 0. 
We see that the exact 
simulation result practically coincides with the results obtained via the 
transfer matrix approach. In these cases the monomer density has a finite 
value at the wall. In the case of the Schr\"odinger approach, where 
the chain continuity error is present, the radial monomer density 
vanishes at the wall, as expected. 

Since above we have considered a Gaussian chain without excluded volume
effects, there is no saddle-point error in this calculation; we are left with 
the lattice discretization and chain continuity errors. From a 
comparison between the 1D and 3D transfer matrix results in Fig. 2, it is 
clear that in this case the lattice discretization error is practically 
non-existent, as the results from the fine 1D lattice virtually coincide 
with these from the rough 3D lattice. From the same figure it is also 
clear that the major error in this case of no excluded volume 
interactions is the chain continuity error, which is also insignificant.

\begin{figure}[!]
\vspace{-0.6 cm}
\psfig{file=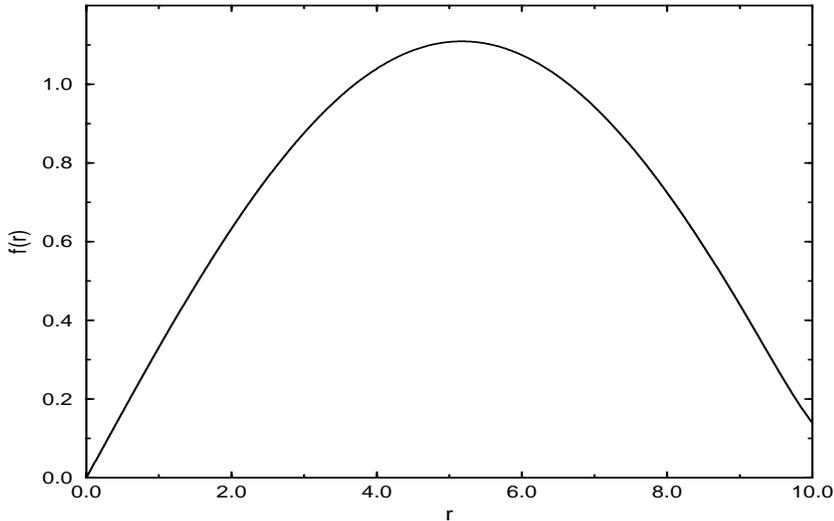,width=370pt,height=240pt,angle=270}
\vspace{-0.4 cm}
\caption{The radial function $f(r)$ exhibiting a finite value at the edge 
of the sphere.}
\end{figure}

\begin{figure}[!]
\vspace{-1.0 cm}
\psfig{file=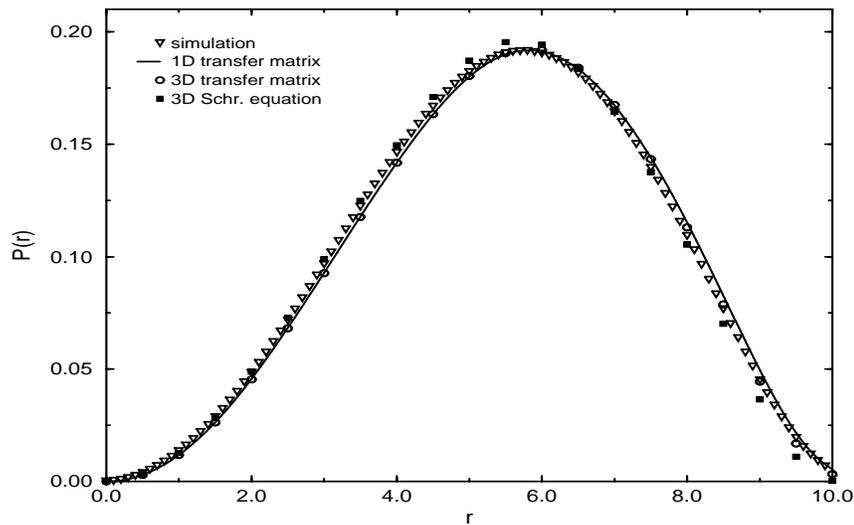,width=370pt,height=240pt,angle=270}
\caption{A comparison between results from the Monte Carlo heat bath 
simulation (exact), the transfer matrix approach, and the Schr\"odinger 
equation approach, for the radial monomer density $P(r)$ of a Gaussian 
polymer chain of 100 monomers without excluded volume interactions 
with a Kuhn length unity in a spherical cavity of radius ten. (The error bars 
of the simulation curve are within the size of the symbols used to 
mark it.) The ``wall effect'' is less apparent in this figure as we are 
plotting the monomer density (roughly speaking, the square of $f(r)$ in 
Fig. 1).}
\end{figure}

In Fig. 3 we show an analogous plot of the radial monomer density with all 
parameters the same as in Fig. 2 but with
excluded volume interactions present with $\lambda=0.0012$. It is clear that 
in this case there is an error due to the saddle-point approximation, leading 
to a deviation between the transfer matrix and the simulation result. Here, 
again, the ``wall effect'' is present in the simulation and the transfer 
matrix density, while the density resulting from the Schr\"odinger equation 
approach vanishes at the wall.

\begin{figure}[!]
\psfig{file=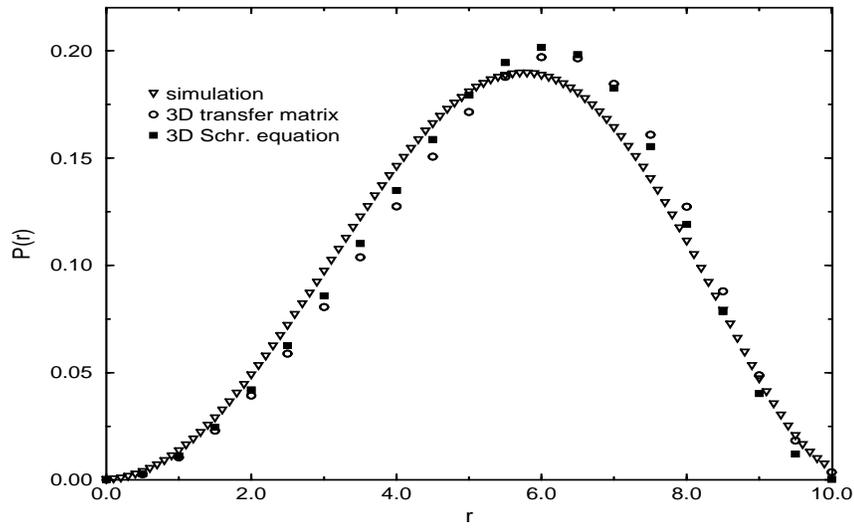,width=370pt,height=240pt,angle=270}
\caption{A comparison between results from the Monte Carlo heat bath 
simulation (exact), the transfer matrix approach, and the Schr\"odinger 
equation approach, for the radial monomer density $P(r)$ of a
polymer chain of 100 monomers with excluded volume interactions with 
$\lambda=0.0012$.  
The Kuhn length is unity and the spherical cavity has a radius of ten. (The 
error bars of the simulation curve are within the size of the symbols 
used to mark it.)} 
\end{figure}

\newpage
\section{Comparison between lattice field theory and simulation results 
for a polymer chain partitioned between two connected spherical 
cavities of different radii}

 The simulation of the system (\ref{Z}) by Monte Carlo techniques is 
readily accomplished
 given the absence of long-range forces. We have chosen to perform the 
simulation by
 generating new polymer configurations using a heat-bath algorithm on the 
Gaussian
 part of the Boltzmann measure 
$e^{-\frac{3}{2a_{p}^{2}}\sum_{s=0}^{M-1}(\vec{x}_{s+1}-\vec{x}_{s})^{2}}$, 
rejecting the obtained configuration if it is forbidden by the
 exclusion potential $V_{\rm excl}$ (here assumed to be either zero or 
$\infty$), and then correcting for the remaining Yukawa  potential
 with a Metropolis accept/reject step. Specifically, the algorithm employed 
is as follows:
\begin{enumerate}
\item  First, a monomer location $i$ is chosen at random 
(i.e. an integer between 0 and $M-1$)
 and a new location $\vec{x}_{i}$ generated according to the weight 
\begin{equation}
e^{-\frac{3}{2a_{p}^{2}}\sum_{s=0}^{M-1}(\vec{x}_{s+1}-
\vec{x}_{s})^{2}} \propto e^{-3(\vec{x}_{i}-\vec{d}_{i})^{2}/a_{p}^{2}},
\label{weight}
\end{equation}
 where $\vec{d}_{i}\equiv \frac{1}{2}(\vec{x}_{i-1}+\vec{x}_{i+1})$.
\item If the new location $\vec{x}_{i}$ is outside the allowed region 
(as determined by the exclusion potential) it is rejected and a new 
$\vec{x}_{i}$ found by the preceding algorithm.
\item The Yukawa part of the Hamiltonian, 
$S_{\rm Yuk}\equiv\frac{\lambda}{2}\sum_{s,s^{\prime}}V_{\rm Yuk}(\vec{x}_{s}-\vec{x}_{s^{\prime}})$
 is used to accept/reject the new configuration (Metropolis update), 
with Yukawa Hamiltonian $S_{\rm Yuk}^{\prime}$ as follows:\\
(a) if $S_{\rm Yuk}^{\prime}<S_{\rm Yuk}$, the new configuration is accepted 
without  further ado,\\
(b) if $S_{\rm Yuk}^{\prime}>S_{\rm Yuk}$, the new configuration is accepted 
with probability $e^{-S_{\rm Yuk}^{\prime}+S_{\rm Yuk}}$. If it is rejected,
the monomer is left unmoved.
\end{enumerate}

In order to obtain confidence in our results, we have performed a 
second dynamic 
Monte Carlo simulation based on the kink-jump technique developed by 
Baumg\"artner and Binder \cite{Baum}. In this method the polymer chain is 
treated as a ``pearl-necklace'' in which every two consecutive monomers 
are connected by a rigid rod of length $l$. Monomers are modeled as hard 
spheres, whose radius $r_{m}$ can be adjusted to describe the strength of 
the excluded volume interaction. Initially the chain is placed in the allowed 
region with the positions of its monomers chosen randomly. Then the chain 
dynamics is evolved by the kink-jump technique: at each time step a monomer 
is picked, say the $n$th one, and then it is rotated around the axis 
connecting the 
$(n-1)$th and the $(n+1)$th monomers at a random angle. If an end monomer 
is chosen, it is moved to a new random position keeping the rod length 
 between it and the monomer to which it is connected fixed. 
The move is accepted if the monomers do not overlap and stay within the 
allowed region. 

We apply these two simulations methods to a polymer chain confined to move 
within two connected spheres of different radii. In relative units, we 
choose the big sphere to have a radius $R_{1}$ equal to unity, the small one 
has a radius $R_{2}=0.8$, the Kuhn length is chosen to be $a_{p}=0.2$, and 
the spheres are embedded in each other with the distance between their 
centers fixed at 1.7. This system is schematically shown in Fig. 4. 

\begin{figure}[!]
\hspace{1.2 cm}
\psfig{file=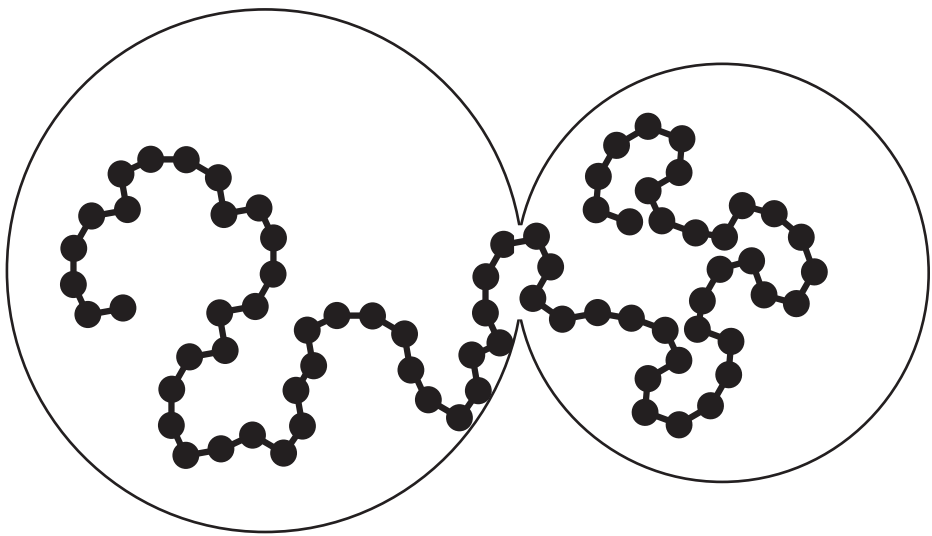,width=350pt,height=200pt,angle=0}
\caption{A polymer chain confined to move within two connected spheres.}
\end{figure}

We have calculated the polymer partition coefficient $K$, defined 
as the ratio of the average number of monomers in the big and small spheres 
respectively, $K\equiv\left<M_{1}\right>/\left<M_{2}\right>$, where the 
total number of monomers is $M=\left<M_{1}\right>+\left<M_{2}\right>$, for 
varying $M$. First, in Fig. 5 we show the simulation results for the case of
a non-interacting chain. These results are compared with lattice field 
calculations using the transfer matrix and the Schr\"odinger equation 
approaches. As in the case described in the previous section, we see that 
the agreement between all the different approaches is very good for the case 
of zero interaction between the monomers, which confirms once again that 
the chain continuity and lattice discretization errors are small. In this 
figure the result from the Schr\"odinger equation approach appears to be
closer to the statistically exact simulation result than the transfer 
matrix result, probably due to 
a coincidental cancellation of errors. Here, and in the next two figures, we
observe that in the case of longer chains the transfer matrix approach 
produces a partition coefficient 
which is systematically smaller than the one resulting from the simulations.

\begin{figure}[!]
\psfig{file=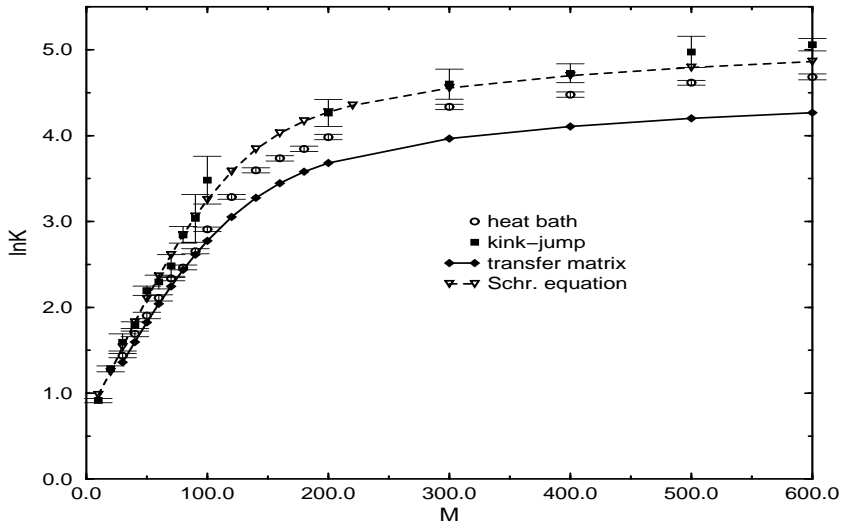,width=370pt,height=240pt,angle=270}
\caption{$\ln{K}$ vs $M$ for the case of a Gaussian chain with no 
excluded volume interactions. The 
simulation results are seen to agree with the lattice field calculations.}
\end{figure}

In Fig. 6 we show results for $\ln{K}$ 
from the simulations for the interacting chain 
with $\lambda_{1}=0.0012$ and $\gamma=a_{p}^{2}$ for system (\ref{Z}), 
which is matched to a monomer radius 
$r_{m,1}=0.025$ in the kink-jump technique. These results are also compared 
to the results obtained via the transfer matrix lattice field calculation 
for the same parameters used in the Monte Carlo heat bath/metropolis 
calculation. In this calculation the two-sphere system is modeled on a 
discrete cubic lattice of 44 points on each side with lattice spacing 
$a=0.1$ as described in \cite{Ts, TCD}. 
In our previous work \cite{Ts} we have shown that 
the size of the aperture connecting the two spheres plays an important role
in determining the value of the partition coefficient $K$. However, when 
the system shown in Fig. 4 is modeled on a lattice, lattice artifacts 
prevent us from obtaining the exact same aperture size, as in the 
continuous case. Therefore, the 
aperture size on the lattice is adjusted to be as close as possible to the 
one in the continuous case shown in Fig. 4. 

In our previous work \cite{TCD2} we have shown that the excluded volume 
parameter 
$\lambda$ can be connected to the monomer radius via the second virial 
coefficient, leading to the expectation that
\begin{equation} 
\lambda\approx{\rm const}\times r_{m}^3.
\label{virc}
\end{equation}
 From the parameters of the calculation 
whose results are presented in Fig. 6 we calculate ${\rm const}$ and use it 
to calculate a second excluded volume 
parameter, $\lambda_{2}=0.0033$, corresponding to the 
monomer radius $r_{m}=0.035$, and then we perform 
simulations for the two new sets of parameters by the heat bath/metropolis 
method and the kink-jump technique, and present the results in Fig. 7. 
The good agreement between this second set of simulations 
confirms formula (\ref{virc}).  
We have also performed the mean-field transfer matrix calculations for this 
new $\lambda$,
 and have plotted the results for comparison in the same figure. 

\begin{figure}[!]
\psfig{file=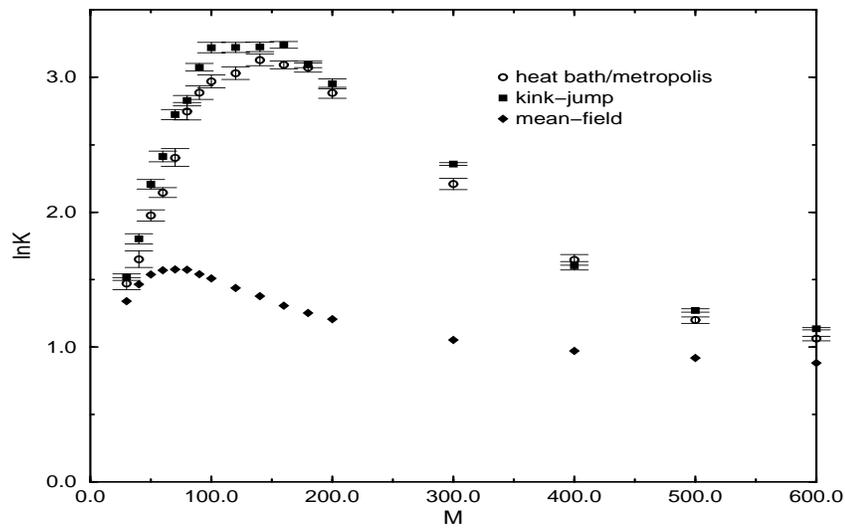,width=370pt,height=240pt,angle=270}
\caption{$\ln{K}$ vs $M$ for $\lambda=0.0012$, which is matched to  
$r_{m}=0.025$. The mean-field transfer matrix results for the same $\lambda$ 
are seen to disagree substantially with the statistically exact results, 
although in both curves there is a maximum with a turnover.}
\end{figure}

\begin{figure}[!]
\psfig{file=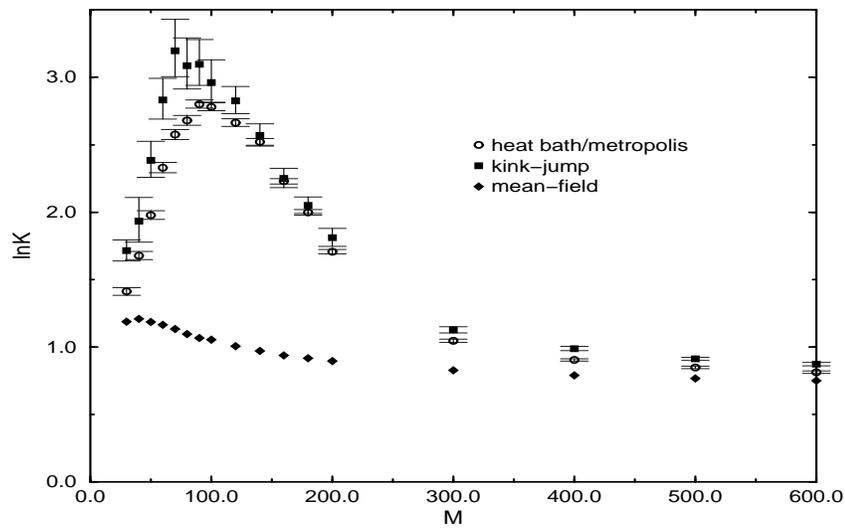,width=370pt,height=240pt,angle=270}
\caption{$\ln{K}$ vs $M$ for $\lambda=0.0032928$, corresponding to 
$r_{m}=0.035$ in the kink-jump simulation technique. The mean-field 
transfer matrix results for the same $\lambda$ are seen to disagree 
with the simulation results, although both curves possess the same generic 
features.}
\end{figure}

From Figs. 6 and 7 we observe that $\ln{K}$ increases almost linearly 
with $M$ for small $M$, then goes through a turnover, and at large $M$ 
decreases to a 
limiting value determined by the ratio of the volumes of the two spheres. 
From the simulations we observe that, as is to be expected, the 
turnover occurs when the end-to-end distance of the polymer chain becomes 
comparable to the diameter of the larger sphere. Both Figs. 6 and 7 
 show that although the mean-field results show the same generic 
behavior, they deviate substantially from the exact simulation results. 
Since we have eliminated the chain continuity error by using the 
transfer matrix approach, and we have already seen that the lattice 
discretization error in these calculations is relatively small, we conclude 
that the quantitative discrepancy between the statistically exact 
simulation results and the mean-field results is due mainly to the 
saddle-point approximation implicit in the mean-field approach. Also, it 
should be noted that in these cases the mean-field results from the 
Schr\"odinger equation approach (not shown in the figures) are very close to 
the transfer matrix results, leading to the conclusion that the 
chain continuity error is small in the case of excluded volume interactions 
too.
 
The results show that the error is larger in the cases of shorter  
chains, while in the cases of longer chains it is relatively small. This 
can be explained with the help of the following observation made in the 
course of the  
simulations: the shorter chains tend to be localized entirely in one 
sphere or the other (thus achieving
a higher conformational entropy), and only make 
infrequent jumps between the two spheres. 
This creates a highly fluctuating monomer repulsion field and renders 
the saddle-point approximation less reliable. In the case of a long chain, 
monomer repulsion competes with the free energy gain due to the higher 
conformational entropy, and the chain tends to stretch between the two 
spheres, occupying both of them 
simultaneously for most of the time. In that case 
the monomer repulsion field exhibits smaller fluctuations and the 
saddle-point approximation is expected to be better. To illustrate 
this point, in Fig. 8 we show a plot of the 
number of monomers in the bigger sphere, $M_{1}$, as a function of the number 
of simulation sweeps, for two different cases, $M=500$ and $M=100$, for 
the parameters corresponding to Fig. 7.  
It is seen that the relative fluctuations in $M_{1}$ are much larger for the 
smaller $M$. 

\begin{figure}[!]
\hspace{-6.35 in}
\psfig{file=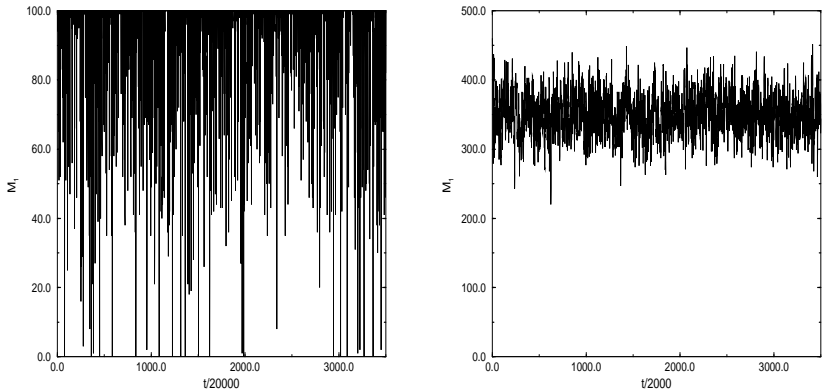,width=880pt,height=850pt,angle=270}
\vspace{-7.9 in}
\caption{The number of monomers in the bigger sphere, $M_{1}$,
as a function of the number of simulation sweeps $t$ for $M=100$ (left plot) 
and $M=500$ (right plot), for the parameters corresponding to Fig. 7.}
\end{figure}

It is also important to point out that what we actually measure in the 
simulations is the average number of monomers in the big sphere, 
$\left<M_{1}\right>$, and the 
partition coefficient $K$ computed from it is very sensitive to small 
errors in $\left<M_{1}\right>$, as $\left<M_{1}\right>$ and 
$\left<M_{2}\right>$ are constrained to add to $M$, hence a small error in 
$\left<M_{1}\right>$ will automatically be reflected in 
$\left<M_{2}\right>$ with an opposite sign, thus amplifying the error in the  
ratio $K$, especially when $\left<M_{1}\right>$ is much bigger than 
$\left<M_{2}\right>$, as in the calculations presented here. 
The actual error between the mean-field 
calculations and the simulations is, in fact, much smaller than one 
would infer from Figs. 6 and 7. To illustrate this point, 
in Fig. 9 we plot the results 
for $\left<M_{1}\right>$ and $\left<M_{2}\right>$ as a function of $M$ 
from the calculations corresponding to the $\ln{K}$ plot in Fig. 7. 

\begin{figure}[!]
\hspace{-6.35 in}
\psfig{file=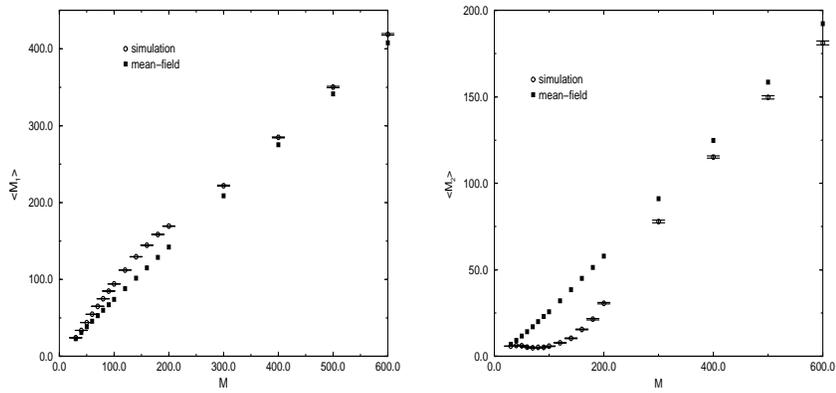,width=880pt,height=850pt,angle=270}
\vspace{-7.9 in}
\caption{The average number of monomers in the bigger and smaller spheres, 
$\left<M_{1}\right>$ (left plot) and $\left<M_{2}\right>$ (right plot), 
respectively, as a function of the total 
number of monomers, $M$, for the parameters corresponding to Fig. 7.}
\end{figure}

\newpage
\section{Conclusions}

The main focus of this work has been investigation of the 
validity and reliability of mean-field methods 
in polymer statistical mechanics by systematic elimination and estimation of
all errors inherent to the lattice field theory approach developed in our
earlier work \cite{TCD2,TCD},
apart from the saddle-point approximation implicit in the 
mean-field method. To be able to evaluate these errors we have compared 
the lattice field theory results 
to the statistically exact results from two independent 
Monte Carlo simulations, namely, a heat bath/metropolis computation and 
a computation based on the kink-jump simulation technique. In the case of 
no excluded volume interactions, computation of the radial monomer density 
of a polymer chain constrained to move within a spherical cavity, and of 
the partition coefficient of a chain restricted to move within 
two connected spheres, shows that ``chain continuity'' 
and ``lattice discretization errors'' are relatively small. On 
the other hand, including short-range pair-wise interactions in the 
Hamiltonian  
and subsequent mean-field treatment of these via the saddle-point 
approximation leads to a larger quantitative error in the partition 
coefficient for problems of certain geometry, as is shown by a comparison 
with the exact simulation results.

This error is larger for shorter chains, and diminishes for long chains. 
It has been attributed to the strongly fluctuating monomer repulsion 
field in the case of short chains, as revealed by the simulations: the 
short chains preferentially localize entirely in one sphere or the 
other (thus attaining a higher conformational entropy),
 making infrequent jumps between the two spheres, which leads to a strongly 
fluctuating repulsion field felt by each monomer in the chain, 
and diminishes the 
validity of the saddle-point approximation, an approximation which assumes 
smoothly varying 
fields. In the case of long chains, the large number of monomers leads to 
a very strong repulsion, overwhelming the gain in conformational 
entropy which would have been achieved had the whole chain been localized 
in one of the two spheres, hence 
the chain stretches into both spheres in its most likely configuration, 
creating a smoother interaction field felt by the individual monomers. Thus, 
in this case the saddle-point approximation is expected to be good, and 
indeed, the mean-field and the simulation results show good agreement.

The calculations presented in this paper suggest that the mean-field approach 
works very well for problems of polymers moving easily in an open region, 
but is not very accurate in situations where the polymers are forced to move 
between several such regions connected only by narrow conduits leading 
to large fluctuations in the monomer density field. Fortunately, in certain 
cases it turns out to be possible to ``weld'' the mean-field results for the 
polymer partition function in separate open regions into an accurate 
description of the more complicated situation, so that the saddle-point 
approximation is only involved where it is accurate. In this way we have 
been able to reproduce the simulation results of Figs. 6 and 7 for 
$\ln{K}$. A detailed description of this technique will be presented in a 
forthcoming publication.
\\

{\bf Acknowledgment}: This work was supported by the National Science 
Foundation Grant CHE-9633561. Some of the computations presented were 
performed at the University of Pittsburgh's Center for Molecular and 
Materials Simulation. The work of A.D. was supported in part by NSF grant 
97-22097.

\end{document}